\documentclass[10pt,conference]{IEEEtran}

\AtBeginEnvironment{tabular}{\footnotesize}
\IEEEoverridecommandlockouts

\AtBeginDocument{%
  \providecommand\BibTeX{{%
    \normalfont B\kern-0.5em{\scshape i\kern-0.25em b}\kern-0.8em\TeX}}}

\newif\ifdraft
\drafttrue
\usepackage[numbers,sort&compress,square]{natbib}
\usepackage{soul}
\usepackage{graphicx}
\usepackage{xurl}
\usepackage{booktabs}
\usepackage{multibib}
\usepackage{natbib}
\usepackage{hhline}
\usepackage{xcolor}
\usepackage{graphicx}
\usepackage{colortbl}
\usepackage{multirow}
\usepackage{tabularx}
\usepackage{amsmath}
\usepackage{makecell}
\usepackage{textgreek}
\usepackage{tabularx}
\usepackage{siunitx}
\usepackage{stfloats, caption}%
\usepackage{tablefootnote}
\newcolumntype{Y}{***{\centering\arraybackslash}X}
\usepackage{bookmark}
\usepackage{textcomp}
\usepackage{xcolor}
\usepackage{microtype}
\usepackage{booktabs}
\usepackage{etoolbox}
\usepackage{enumitem} 
\usepackage{url}
\usepackage{ragged2e}
\usepackage{amsmath}
\usepackage{booktabs}
\usepackage{multirow}
\usepackage{tabularx}
\usepackage{svg}
\usepackage{mdframed}

\author{\IEEEauthorblockN{Template}}

\newcolumntype{P}[1]{***{\raggedright\let\newline\\\arraybackslash\hspace{0pt}}m{#1}}

\newcolumntype{C}[1]{***{\centering\let\newline\\\arraybackslash\hspace{0pt}}m{#1}}

\newcommand{\MyBox}[1]{%
\vspace{3mm}
\begin{mdframed}
#1
\end{mdframed}
\vspace{3mm}
}

\begin{document}

%\title{Evaluation of Cognitive Workload During Code Comprehension Tasks\\
\title{Feeling Easier, Working Harder: Cognitive Load in AI-Assisted Code Comprehension}

%\author{\IEEEauthorblockN{Anonymous}}

\author{\IEEEauthorblockN{Maryangela White}
\IEEEauthorblockA{\textit{Computer Science Department} \\
\textit{Colorado State University}\\
Fort Collins, United States \\
maryangela.white@colostate.edu}

\and

\IEEEauthorblockN{Maximilian Rose}
\IEEEauthorblockA{\textit{Computer Science Department} \\
\textit{Colorado State University}\\
Fort Collins, United States \\
maximilian.rose@colostate.edu}

\and

\IEEEauthorblockN{Matthew Buckman}
\IEEEauthorblockA{\textit{Computer Science Department} \\
\textit{Colorado State University}\\
Fort Collins, United States \\
matthew.buckman@colostate.edu}

\and 
\IEEEauthorblockN{Amy McDonnell} 
\IEEEauthorblockA{\textit{Psychology Department} \\
\textit{Colorado State University} \\
Fort Collins, United States \\
amy.mcdonnell@colostate.edu}

\and 

\IEEEauthorblockN{Francisco Ortega} 
\IEEEauthorblockA{\textit{Computer Science Department} \\   
\textit{Colorado State University} \\ Fort Collins, United States \\
f.ortega@colostate.edu}

\and

\IEEEauthorblockN{Bianca Trinkenreich} 
\IEEEauthorblockA{\textit{Computer Science Department} \\
\textit{Colorado State University} \\ Fort Collins, United States \\
bianca.trinkenreich@colostate.edu}

}

\maketitle

\begin{abstract}
\textbf{Background}: 
Software developers spend significant time reading and understanding code, a task that is increasingly being supplemented by generative AI. While these tools aim to reduce the intrinsic difficulty of reading code, interacting with them introduces a dual-task environment that may impose extraneous cognitive load that is challenging to capture.\\
%from traditional subjective surveys.
\textbf{Aims}: This study investigates how access to generative AI affects developers' real-time cognitive workload (RQ1), objective code comprehension performance (RQ2), and self-perceived code comprehension strategies (RQ3).\\
\textbf{Method}: We conducted a 3x2 within-subjects study with 20 novice computer science students who reviewed Java code snippets of varying lengths with or without AI assistance, followed by an AI-unassisted comprehension quiz. Objective cognitive workload was continuously captured using a standardized tactile Detection-Response Task (tDRT) during both reading and quiz phases, and then compared to subjective post-quiz NASA-TLX and SEQ scores.\\
\textbf{Results}: AI assistance significantly increased objective cognitive workload, raising tDRT reaction times by 11.2\%, yet participants' perceived workload remained unchanged.  Code comprehension scores were high under both conditions, with no evidence that AI assistance improved comprehension at the difficulty levels evaluated; scores clustering near the maximum limited power to detect benefits on harder tasks. Qualitatively, participants reported trusting the AI conditionally---verifying its output against their own reading for more complex snippets, but directly trusting it for simpler code---and, while many felt their reading process was unchanged, a notable subset shifted toward skimming or reading the AI's explanations in place of the code itself.\\
\textbf{Conclusions}: 
These findings reveal that developers do not accurately perceive the cognitive resources they expend when managing AI tools. This study provides evidence of added cognitive effort associated with using generative AI, pointing towards a need to account for this dual-task penalty in software engineering tools and education.
\end{abstract}


\section{Introduction}
\label{sec:introduction}

For decades, software developers have spent up to 70\% of their time reading and understanding code rather than writing it~\cite{xia2017measuring, minelli2015know}. Generative AI (GenAI) assistants promise to ease this burden by translating convoluted logic into clear, natural-language explanations~\cite{nam_using_2024}. However, emerging evidence suggests that human-AI collaboration splits the developer's attention, trading the effort of writing code for the hidden mental strain of prompt-formulation, context-switching, and output verification~\cite{mozannar_reading_2024, tang_developer_2024}.
Because this mental effort is largely hidden behind developers' generally positive impressions of GenAI tools, they frequently underestimate the cognitive effort they actually expend~\cite{vaithilingam2022expectation,yu_efficiency-gain_2026}. This means traditional cognitive workload measurements like the NASA Task Load Index~\cite{doi:10.1177/154193120605000909} fail to capture this discrepancy, as post-task surveys only capture what a developer remembers after the task is over~\cite{yeh1988dissociation}.  

% Standard instruments such as the NASA Task Load Index~\cite{doi:10.1177/154193120605000909} and SEQ scale~\cite{sauro_chapter_2012} capture retrospective self-perceptions, failing to isolate this critical cognitive dissociation. 
%The consequences of this dual-task environment are beginning to surface. Developers have been observed outsourcing comprehension without adequate verification~\cite{ross2023programmer}, over-reliance on AI output has been linked to automation bias and skill degradation~\cite{kazemitabaar_exploring_2025}, and novice developers in particular exhibit distinct over-reliance patterns such as ``shepherding'' and ``drifting''~\cite{prather2024widening}. 
%Even when developers engage critically, verifying AI output carries its own measurable mental cost~\cite{akhoroz2025conversational}. The problem is that this cost is largely invisible, because the general perception of GenAI is favorable~\cite{vaithilingam2022expectation}, developers are likely to underestimate the cognitive effort they actually expend~\cite{yeh1988dissociation, fritz2014using}. 
%what developers \textit{think} they experienced, not what their cognitive system actually expended~\cite{yeh1988dissociation, fritz2014using}, a dissociation that relying on self-report alone cannot resolve.

To uncover this hidden mental workload, researchers have commonly relied on continuous physiological tracking (e.g. EEG, fNIR, and eye-tracking)~\cite{obaidellah2018survey}.
While these methods yield rich data, they require expensive equipment, fragile calibrations, and noise-sensitive signals, making them impractical for everyday software engineering environments. 
We address this instrumentation gap by introducing the Tactile Detection-Response Task (tDRT) to SE research. The tDRT is a low-cost, ISO-standardized behavioral instrument~\cite{noauthor_iso_nodate} that quantifies cognitive workload via fluctuations in reaction times to simple physical stimuli.

%studies, leaving the field without a viable objective measure for everyday use.
%to map cognitive resource allocation during SE tasks such as reading and writing code~\cite{gonccales2021measuring, obaidellah2018survey, krueger2020neurological, sanchez2018artificial}. 
 %demonstrated that complexity metrics only partially explain the neural correlates of program comprehension~\cite{peitek_program_2021}. 
%alternative for measuring real-time cognitive workload in code comprehension studies. The tDRT is an ISO-standardized human factors instrument~\cite{noauthor_iso_nodate} that quantifies cognitive workload via fluctuations in reaction times. Beyond its extensive validation in automotive safety research, the tDRT has been successfully adapted to measure cognitive workload across broader human-computer interactions (HCI)~\cite{innes_broader_2021,thorpe_assessing_2019} and extended reality environments~\cite{maag_measuring_2023}. To the best of our knowledge, this study represents the first application of this methodology within software engineering research. 

By leveraging the real-time tracking capabilities of the tDRT alongside traditional performance and subjective metrics, we can better examine the hidden costs of AI-assisted program comprehension. To guide this investigation, we present three core research questions:

\vspace{2mm}
\noindent\hangindent0.5em\textit{\textbf{RQ1.} How does access to GenAI assistance affect developers' real-time cognitive workload during code comprehension?}

%Understanding this cognitive cost is essential for developers, educators, and tool designers seeking to make informed decisions about when and how to deploy AI assistance. Without objective measurement, developers risk operating under the false impression that AI is reducing their mental effort, when they may in fact be accumulating undetected cognitive fatigue, a risk compounded by the finding that subjective workload ratings do not reliably reflect objective cognitive effort~\cite{yeh1988dissociation, fritz2014using}.

\vspace{2mm}
\noindent\hangindent0.5em\textit{\textbf{RQ2.} How does access to GenAI assistance affect observable code comprehension performance across tasks of varying complexity?}

%Quantifying the cognitive cost of AI tools does not tell us whether that overhead translates to a tangible benefit. A higher workload may be an acceptable trade-off if it yields better comprehension outcomes. Conversely, it becomes counterproductive if performance stalls, particularly for novice developers still building foundational mental models of software architecture~\cite{sweller_cognitive_2011, prather2024widening}. If AI assistance does not improve comprehension outcomes despite its cognitive overhead, students who rely on it may be investing mental effort for no educational return. This leaves them poorly equipped for situations where independent performance is required, such as technical interviews, professional code reviews, or academic assessments)~\cite{prather2024widening}.

\vspace{2mm}
\noindent\hangindent0.5em\textit{\textbf{RQ3.} How do developers perceive the impact of GenAI assistance on their overall code comprehension process and strategies?}

%While quantitative measures can help us evaluate whether AI changes workload or performance, they cannot capture how developers adapt their cognitive strategies when using an AI tool. GenAI usage %or Human- AI interaction requires developers to actively navigate the dilemma of trust, balancing the speed of AI responses against the risk of incorrect or misleading information. If developers must continuously verify AI output, their code-reading strategies may fundamentaxlly shift from traditional linear analysis~\cite{busjahn_eye_2015} to cross-examination and vetting~\cite{mozannar_reading_2024}. Exploring these self-perceived changes in strategy and trust is essential for understanding the behavioral patterns that drive our quantitative findings.  

To explore these questions, we conducted a $3 \times 2$ within-subjects experiment with 20 novice computer science students. Participants evaluated Java code snippets across three sizes under both AI-assisted and unassisted conditions. Cognitive workload was captured continuously using the tDRT and supplemented by post-task NASA-TLX and SEQ metrics. Comprehension performance was assessed via multiple-choice questions probing maintenance, summarization, mental execution, and structural understanding~\cite{10.1145/3626522},  question types that mirror realistic code review demands in collaborative software projects~\cite{jain_2019_codesnippets, 7961505}. Finally, to capture changes in strategy, we conducted a qualitative analysis of participant experiences, focusing on how utilizing the AI altered their reading methods and calibration of trust. 

Our results show that AI assistance significantly \textit{increased} objective cognitive workload (RQ1), raising tDRT reaction times by 11.2\% over the unassisted condition ($\chi^2 = 17.37$, $p < .001$), even though subjective questionnaire ratings remained unchanged. Notably, this cognitive load increase persisted even after the AI was no longer in use; during quiz phases (where AI was not accessible), participants in the AI-condition still exhibited reaction times $6.93\%$ higher than the control group ($p=0.007$).
%This increase in reaction time relative to the control group also varied across code snippet sizes, with the medium-sized snippets showing the largest gap between conditions, however this difference was due to a dip in reaction time within the control condition for medium snippets, rather than a workload spike in the AI-assisted condition. 
Ultimately, this discrepancy demonstrates that computer science students from our study did not accurately perceive the cognitive resources expended when using GenAI for code comprehension. The subjective ease of interacting with GenAI may mask a measurable, sustained objective cost. 
%a ceiling effect prevented a definitive conclusion, though exploratory analyses of sub-maximal scores suggest 
For code comprehension performance (RQ2), scores were high under both conditions, and we found no evidence that AI assistance improved comprehension at the difficulty levels evaluated. 
Qualitatively (RQ3), developers reported adapting their strategies based on task difficulty, where they verified AI output against their own reading for complex snippets, but directly accepted explanations for simpler code. Furthermore, while many participants felt their habits remained unchanged between conditions, a notable subset shifted towards skimming the code during the AI-assisted condition, or relying entirely on the AI's natural language explanation.

This paper makes novel contributions to SE research by introducing the tDRT methodology within an SE context and providing empirical evidence of the hidden cognitive costs that accompany generative AI usage in code comprehension. The remainder of this paper is structured as follows. Section~\ref{sec:related_work} reviews existing research on cognitive workload and generative AI usage in software engineering. Sec.~\ref{sec:researchdesign} details our data collection and analysis, followed by results in Sec.~\ref{sec:results}, and discussion in Sec.~\ref{sec:discussion}. Finally Sec.~\ref{sec:conclusion} concludes the paper and provides suggestions for possible directions of future work.

\section{Related Work}
\label{sec:related_work}

\subsection{LLM Use in Software Engineering}

The integration of Large Language Models (LLMs) into software development has radically altered the cognitive landscape of code comprehension. As code synthesis tools become widespread, a developer's primary task frequently shifts from active code generation to passive code reading, critical evaluation, and verification~\cite{qiao_systematic_2026, haque_towards_2025}. Neural imaging research confirm that the cognitive profiles of reading and writing code are neurologically distinct~\cite{krueger2020neurological}, meaning that developers adopting a ``reviewer'' role alters their underlying mental demands. Rather than outright reducing mental effort, outsourcing code generation to an AI assistant trades the cognitive load of syntax synthesis for the load of structural analysis. 

This transition introduces a compounding behavioral cycle: because analyzing unfamiliar, AI-generated code remains mentally taxing, developers frequently escalate their reliance by offloading the comprehension process itself. Rather than analyzing the code independently, developers employ scaffolding tools such as CodeMap or GILT to automate the interpretation of program flow and logic structures~\cite{nam_using_2024, gao2025understanding}. In extreme cases, developers outsource comprehension completely, prompting models to make changes within systems they do not fully understand~\cite{qiao_systematic_2026, nam_using_2024}.
Additionally, while these automated tools aim to streamline comprehension, they can inadvertently increase cognitive load. Engaging with AI-generated code explanations forces the developer into a dual-task workflow where they must simultaneously manage the primary task and calibrate their trust in the AI's accuracy~\cite{tang_developer_2024}. This cognitive overhead can incentivize superficial code skimming, causing developers to overlook latent bugs~\cite{perry_users_2023}, while fostering automation bias that risks long-term skill degradation~\cite{kazemitabaar_exploring_2025}. Because these behavioral states are dynamic and can vary across a task, evaluating them requires continuous, real-time tracking of a developer's workload.

%it can also inhibit a developer's level of understanding.In addition developers can place too much trust in the system, causing them to develop automation bias, relying on the tools to perform their tasks, causing their skills to degrade~\cite{kazemitabaar_exploring_2025}.

%Several widely used chat-based large language models are currently available, including ChatGPT, Gemini, and DeepSeek. While limited work has directly compared these systems in code comprehension contexts, prior studies evaluating their performance on programming tasks suggest differences in response behavior and latency. For example, Shahzad et al. found that DeepSeek produced highly accurate solutions, but could produce variable response times~\cite{shahzad_comparative_2025}. Because the code comprehension phase of our study was time-constrained, a system with relatively consistent, fast response latency was preferred to reduce the likelihood that delays would impact participant task completion. Consequently, the publicly available web interface of Gemini was used in the AI-assisted condition. 

\subsection{Measurement of Cognitive Load}
The NASA Task Load Index has been the standard for measuring perceived cognitive load for nearly 40 years~\cite{doi:10.1177/154193120605000909}, however its subjectivity has drawbacks~\cite{yeh1988dissociation, fritz2014using}. For objectively measuring real time cognitive load, physiological modalities such as eye tracking and electroencephalograms (EEGs) can be processed to gauge mental effort. While these devices enable live tracking during a task, they are often cost-prohibitive, require highly technical calibration, and are sensitive to environmental noise~\cite{obaidellah2018survey}. 
To avoid these issues, behavioral modalities such as the Detection Response Task (DRT) offer an alternative, well-validated method for measuring real-time cognitive workload by requiring participants to respond to periodic stimuli while performing a primary task~\cite{stojmenova_detection-response_2018}. These devices have been widely used throughout automotive research to assess distractions while driving. Their application has also begun to expand into other human factor domains, such as software engineering and interface evaluation~\cite{stojmenova_detection-response_2018, innes_broader_2021}. Because the DRT provides a robust, low-overhead method for tracking fluctuations in cognitive workload, we utilize this metric as a core behavioral measure in our study. 

\subsection{Dynamics of Code Comprehension}
% other option for section title: Factors Influencing Code Comprehension
Program comprehension is driven by a dynamic interaction between the developer's internal cognitive strategies and external code characteristics. Literature establishes two primary paradigms of code reading: bottom-up processing, where syntax is mentally grouped into higher-level abstractions, and top-down processing, where existing domain knowledge and semantic ``beacons'' are utilized~\cite{storey2005theories}. Rather than relying on a single mode, developers typically alternate between these strategies based on the demands of the task and available mental resources~\cite{von_mayrhauser_program_1995}. 

Code scale can heavily influence these cognitive shifts: small code snippets facilitate linear, bottom-up comprehension, while larger code pieces encourage a shift towards top-down abstraction~\cite{von_mayrhauser_dynamic_1994}. Consequently, the impact of AI-assistance during comprehension may vary between code sizes. For smaller snippets requiring statement-level processing, AI explanations may reduce the line-by-line interpretive load; for larger snippets where effective navigation and structural chunking become vital~\cite{von_mayrhauser_dynamic_1994, von_mayrhauser_program_1995}, AI may instead assist by surfacing high-level structure. The nature and magnitude of AI’s benefit likely differs across these scales. To capture these potential interaction effects, this study evaluates comprehension across three code sizes.  

Because developers alternate between these strategies, evaluating the impact of AI assistance requires a multi-dimensional assessment framework. As such, the code samples and comprehension questions utilized in this study were developed according to the code comprehension framework by Wyrich et al.~\cite{10.1145/3626522}. According to this framework, code comprehension measurement can be divided into four parts: maintenance, program summarization, mental execution, and structural mapping. Evaluating the impact of AI-assistance across these dimensions ensures that our performance metrics capture the multi-layered nature of program comprehension.

%To evaluate the AI-assisted condition against both “bottom-up”(mental execution) and “top-down” (summarization/maintenance) comprehension processes, we designed questions to cover each of the four subdivisions.

%As code size increases, effective navigation skills become vital for comprehension~\cite{xia2017measuring, minelli2015know}; in such cases, AI assistance may be particularly beneficial in summarizing the code's structure and flow. 
%While using open-source code offers ecological validity, it introduces confounding variables such as domain-specific jargon and inconsistent stylistic choices~\cite{10.1145/2568225.2568252}. 

\section{Research Methodology}\label{sec:researchdesign}

Our study investigated the impact of generative AI-assistance on cognitive workload, objective performance, and user perception across varying code sizes. We manipulated two independent variables: code snippet size (small, medium, large) and AI availability (with/without), utilizing a within-subjects design to control for individual differences in programming ability, reading speed, and prior Java familiarity~\cite{xia2017measuring}. 
%, all of which are known sources of variance in code comprehension

Participants completed six Java comprehension tasks. Within each task, tDRT reaction time was recorded as a proxy for real-time cognitive workload, followed by post-task evaluations using the NASA-TLX and SEQ scales. 
%Integrating these questionnaires alongside psychomotor tracking allowed for a comparison between objective and perceived workload. 
After completing each task, participants were given a one-minute rest period before the next task, to help mitigate cumulative fatigue.
This study was conducted with approval from the Institutional Review Board of Colorado State University.

%Based on the size of the code snippet, participants had a maximum amount of time (4, 5, or 6 minutes) to read the code before receiving the comprehension questions. These questions consisted of 4 multiple-choice questions designed to assess participants’ understanding of the code’s functionality, flow, and structure. This quiz section also had a time limit according to the code snippet size (2, 2.5, or 3 minutes). In the AI-assisted condition, the LLM would be available only during the code-reading. 

\subsection{Participants and Recruitment}
The final analysis sample consisted of $N=20$ software engineering students recruited from two Java-based courses at Colorado State University. To justify our sample size, we evaluated statistical strength using a dual approach based on both a standardized two-condition difference~\cite{trommler2025impact} and a simulation-based sensitivity analysis~\cite{green2016simr}. Following common practice in previous DRT studies, which analyze data with linear mixed-effects models but utilize two-condition difference calculations for sample planning, our design ($N=20$, $\alpha=.05$, 80\% power) corresponds to a minimum detectable standardized effect of approximately $d=.66$, falling between Cohen’s conventional medium ($.50$) and large ($.80$) benchmarks~\cite{cohen1992power}. 

However, because this approximation does not capture the repeated-measures and random-effects structure of our data, we also conducted a SIMR simulation-based sensitivity analysis~\cite{green2016simr} using the mixed-effects structure of our primary model (500 simulations run). For our primary comparison of interest (AI vs no-AI condition), our design achieves $80\%$ power at $\alpha =.05$ for an AI-condition coefficient of $0.083$ on $\log_2$ reaction time, corresponding to a detectable change of approximately 5.9\% in RT ($2^{0.083}-1$). (Note: Due to computational constraints, this sensitivity analysis focuses specifically on the main effect of the condition rather than the subsequent interaction models).

The sample had a gender distribution of $60\%$ men, $30\%$ women, and $10\%$ non-binary.
Full demographic information is presented in Table~\ref{tab:demographics}
%The class involves developing a Java-based application following SE practices.

\begin{table}[htbp]
    \centering
    \caption{Participant Demographic and Experience Summary ($N=20$)}
    \label{tab:demographics}
    \begin{tabular}{llc}
        \toprule
        \textbf{Attribute} & \textbf{Category} & \textbf{Percentage (\%)} \\
        \midrule
        Age & 18--20 & 45\% \\
            & 21--23 & 30\% \\
            & 24--26 & 10\% \\
            & 27--29 & 10\% \\
            & 30+    & 5\% \\
        \midrule
        Java Experience & $<$ 1 year & 5\% \\
                        & 1--2 years & 45\% \\
                        & 3--4 years & 35\% \\
                        & 5+ years   & 15\% \\
        \midrule
        LLM Usage       & Never       & 30\% \\
                        & Rarely      & 20\% \\
                        & Occasionally & 5\% \\
                        & Frequently  & 40\% \\
                        & Very Frequently & 5\% \\
        \bottomrule
        \vspace{-6mm}
    \end{tabular}
\end{table}

\subsection{Tactile Detection Response Task (tDRT)}\label{subsec:DRT_methodology}
The Detection-Response Task (DRT) is a standardized human factors tool for measuring fluctuations in cognitive workload during a primary task~\cite{noauthor_iso_nodate}. The participant is asked to respond to a stimulus presented every 3-5 seconds by pressing a button attached to their finger. The DRT’s ability to measure cognitive workload fluctuations comes from the responses to the stimuli slowing or not occurring when the primary task requires greater cognitive resources~\cite{innes_broader_2021}. 

\begin{figure}
    \centering
    \includegraphics[width=0.8\linewidth]{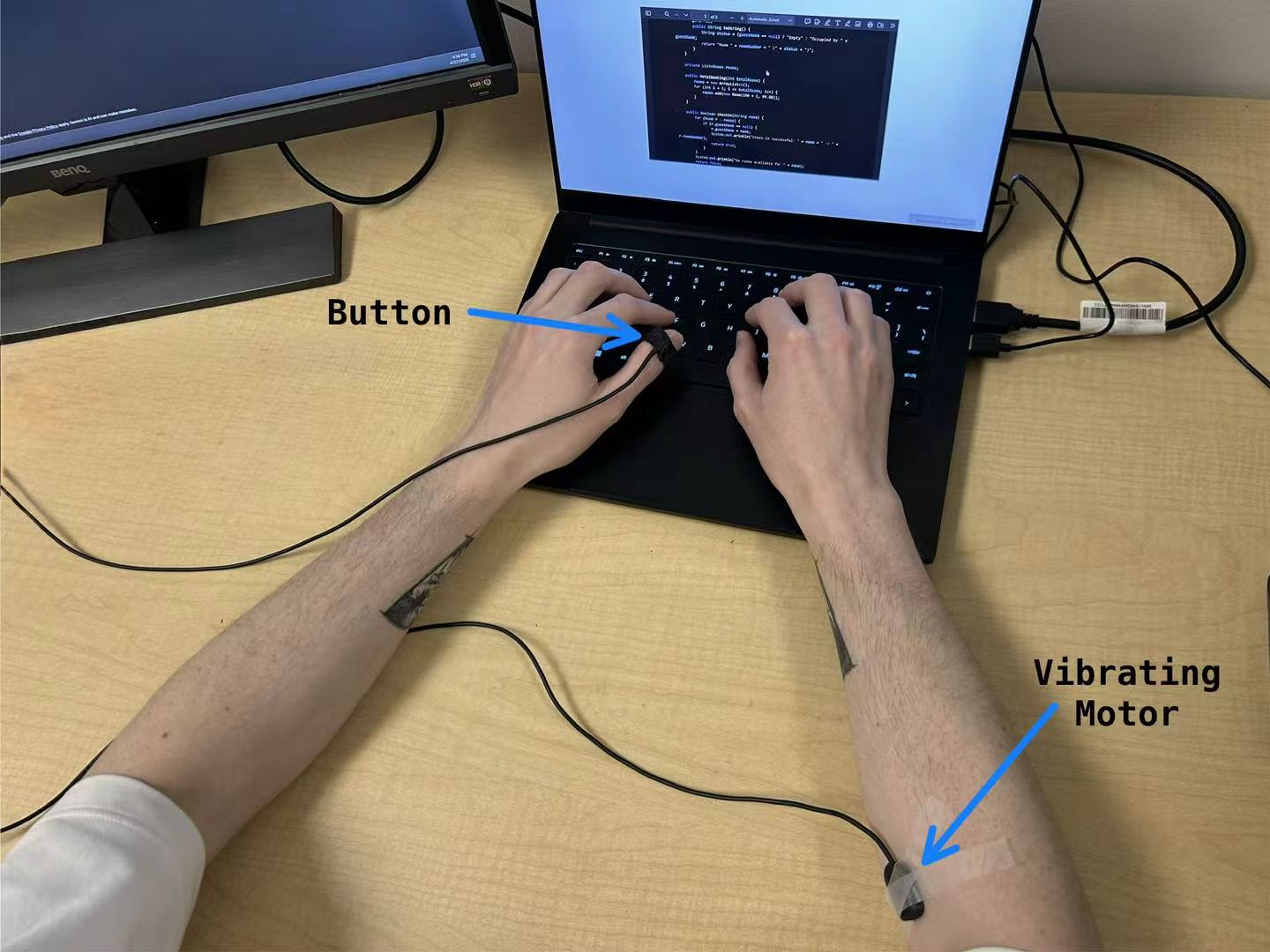}
    \caption{tDRT Set Up}
    \label{fig:DRT_set_up}
\end{figure}

While the DRT supports tactile, visual, and auditory configurations, a tactile stimulus modality was selected to prevent sensory interference with the primary visual code review task~\cite{stojmenova_detection-response_2018}. The stimulus was administered via a commercial tDRT system manufactured by Red Scientific~\cite{phd_detection_nodate}, utilizing a small vibrating motor attached to the participant’s shoulder or arm and a finger-mounted response button (see Figure~\ref{fig:DRT_set_up}).

During the study, tDRT data was recorded across all six tasks and partitioned into reading and quiz segments. To control for confounding physical artifacts, the experimenter manually flagged periods when participants were actively typing, as responding via the finger-mounted DRT button can physically conflict with keyboard use. During pre-processing, these timestamps were cross-checked against screen recordings, and DRT responses falling within those intervals were excluded (similar researcher-recorded temporal markers have been used in prior DRT studies to define task intervals~\cite{cooper_age-related_2020}). Because the goal was simply to identify observable typing periods rather than infer subtle cognitive states, millisecond-level annotation was unnecessary. 
Removing these typing windows, which accounted for 2.6\% of total observations, was a conservative methodological choice. Because typing occurred exclusively during the AI-assisted condition, retaining those confounded trials would have artificially inflated the estimated AI effect (increasing the coefficient from 0.1536 [11.2\%] to 0.1828 [13.5\%]). Furthermore, during the quiz-phase (where no typing occurred in either condition) RT remained significantly higher following AI-assisted tasks. Ultimately, this filtering ensured that the dataset isolated cognitive workload rather than mechanical interference from querying the LLM.

Raw tDRT data underwent standard pre-processing and filtering. In accordance with ISO 17488 guidelines, any response recorded under 100 ms was classified as an accidental trigger and excluded, while signals receiving no response within $2,500$ ms were categorized as misses.~\cite{stojmenova_detection-response_2018}. Only valid responses within this window were included in reaction time calculations, with all misses and invalid triggers omitted. 
Although the DRT yields both reaction time and hit rate metrics~\cite{stojmenova_detection-response_2018}, overall hit rates remained consistently high (above 95\%). This is a common phenomenon in stationary, non-driving DRT applications where participants easily maintain baseline compliance. Because this lack of variance in hit rate restricts statistical modeling, reaction time was selected as the dependent variable for all subsequent DRT analyses. This choice is supported by prior evidence showing that RT remains sensitive even when accuracy metrics show a ceiling effect~\cite{ranney2014detection}. Nevertheless, we conducted a descriptive check: hit rates averaged 93\% in the AI-assisted condition compared to 97\% in the non-AI condition. Although minor, this slight increase in misses under AI assistance aligns directionally with the reaction time findings, offering converging evidence of an increased cognitive load. 
%Following this filtering, a dataset of $8,048$ valid reaction time data points remained for the model analysis 

\subsection{Generative AI Assistance}
During the AI-assisted condition, participants were provided access to the publicly available web interface 
of chat-based Gemini for the code reading phase. Gemini was selected because of the time-constrained nature of the tasks, which required a model with consistently fast response latency~\cite{shahzad_comparative_2025}. To prevent carryover effects between trials, each task used a new chat session opened in an incognito browser window, ensuring that the model had no access to prior conversation history from earlier trials or participants. 

Participants were free to decide how often to consult the AI system during the code review phase. Interaction with the system was free-form; participants could ask questions, paste the provided code snippets, or request explanations, with no constraints on prompt wording or format. No training in prompt engineering was provided to approximate how computer science students typically interact with generative AI tools to understand code. While differences in prompting strategies may influence the type or quality of assistance received, such variability reflects realistic usage patterns and therefore was considered part of the natural interaction with the system. Participants were instructed to avoid inputting any personal or sensitive information. All interactions with the AI system, including prompts and model responses, were captured via screen recording software to assist in typing interval cross-referencing.

\subsection{Code Snippets}

The tasks required participants to review Java code snippets of varying sizes: two small snippets (approx. 30 lines each), two medium-sized snippets (approx. 45 lines each), and two larger code sections (approx. 70 lines each). Manipulating snippet lengths allowed for analysis of code comprehension strategies under different scaling constraints. 

To ensure high internal validity, all code snippets were developed specifically for this study. This allowed for close control over cyclomatic complexity and domain neutrality, ensuring that differences in cognitive workload could be safely attributed to experimental conditions, rather than incidental features of the code~\cite{jain_2019_codesnippets}. Following previous research~\cite{7961505}, snippets isolated specific semantic structures and avoided unfamiliar Java constructs or external library dependencies. Furthermore, to minimize confounding variables such as domain-specific jargon~\cite{10.1145/2568225.2568252}, we followed the Wyrich et al. framework~\cite{10.1145/3626522}, and utilized core language constructs covered within the participants' university curriculum.

\subsection{Code Comprehension Questions}

Code comprehension was assessed via four multiple-choice questions per code task. The questions covered maintenance, summarization, mental execution, and structure \cite{10.1145/3626522} and were designed to require the level of understanding needed to modify or extend the code. This emulates a realistic developer experience where one must comprehend existing logic to contribute to a collaborative project. 

To ensure participants completed their primary analysis during the initial code reading phase, a restrictive time window (2, 2.5, or 3 minutes) was applied to this comprehension phase; the code snippet remained visible during the quiz, but the time constraint discouraged re-reading it in lieu of the prior analysis. Specifically within the AI-assisted tasks, this constraint prevented participants from bypassing the initial code review and simply deferring code reading to the quiz phase when the AI tool was no longer accessible. Complete code snippets and comprehension questions are available in the replication package~\cite{zenodoreppackageDraft}.

\subsection{Experimental Design}
The study employed a $3\times2$ within-subject design, manipulating code size and AI assistance. For each of the three code size categories, two unique code snippets were prepared, one to be completed with AI assistance, and one without. This resulted in six total code review tasks per participant. 

To mitigate potential learning and fatigue effects across the study, the order of the three code size categories was randomized for each participant. 
%For example, one participant might complete the small tasks first, followed by the medium and large code tasks, while another might be presented with the medium-sized tasks first, followed by the small and large tasks. 
Within each size block, the two assigned tasks were completed consecutively. This blocked design was chosen because different code sizes require different reading approaches \cite{busjahn_eye_2015, von_mayrhauser_dynamic_1994}; constantly shifting between snippet lengths could cause frequent context switching that would artificially inflate cognitive load measurements\cite{monsell2003task}. 
While assignment of AI assistance to a code snippet within a size block was counterbalanced across participants, task sequence order within a specific size category was randomized. This within-block randomization prevented order or priming effects without introducing excessive experimental combinations, while also ensuring that each code snippet received an equal allocation of AI-assisted and unassisted trials.

%This within-block randomization was chosen over full counterbalancing to prevent order or priming effects without introducing excessive experimental combinations. This approach kept logistical efficiency while ensuring that each code snippet received an equal distribution of AI-assisted and unassisted trials, preventing comprehension difficulty from confounding evaluation of the conditions. 

\subsection{Questionnaires}
Immediately following each code task, participants completed two questionnaires to measure subjective cognitive load: the NASA-TLX~\cite{doi:10.1177/154193120605000909} and the SEQ~\cite{sauro_chapter_2012}. During this questionnaire window the tDRT device was deactivated. Upon finishing all six tasks, participants completed an exit survey, including demographics (summarized in Table~\ref{tab:demographics}), closed questions about how frequently they use LLM tools for coding and if they have used LLMs for code comprehension before, and open questions about the experiment on the extent to which they trusted the output provided by the LLM during the code comprehension tasks, and whether and how the AI-assistance had changed their approach to reading code. 

%The responses for open questions were qualitatively analyzed for RQ3.

%with representative quotes utilized in Section~\ref{sec:discussion} to contextualize the quantitative results.

\subsection{Data Analysis}
\subsubsection{RQ1 Data Analysis}
To evaluate the impact of generative AI assistance on cognitive workload, tDRT and self-reported outcomes data was analyzed using R (version  $4.6.0$) and the $dplyr$~(v$1.2.1$), $lmerTest$~(v$3.2.1$), $emmeans$~(v$2.0.3$), and $pbkrtest$~(v$0.5.5$) packages.

We used Linear Mixed-Effects (LME) models because they account for the correlation between repeated measurements within participants and accommodate missing data without excluding participants with incomplete observations~\cite{bates_fitting_2015}. Prior to model fitting, the distribution of reaction time data was checked via visual inspection of a QQ-plot. Confirming a characteristic right-skew, reaction time values were $\log_2$-transformed to better meet the assumptions of normality, following established empirical standards~\cite{hammouri_handling_2020}. 

The LME framework splits variables into fixed and random effects. Fixed effects represent the experimental factors manipulated or measured to assess their systematic impact on our dependent variables (reaction time or subjective workload). Importantly, our core fixed effects and the specific interactions we tested -- the main effect of AI-assistance (assisted vs. no-assistance), its interaction with Task Phase (Reading vs. Quiz), its interaction with Code Size (Small, Medium, and Large), and its interaction with Java Experience -- were determined during study design, prior to any data collection or analysis. Java experience was specified as a fixed effect because recruitment was restricted to students with an established minimum proficiency in Java, and we hypothesized that AI-impact would vary by baseline skill. Controlled experimental factors that occurred across all participants, such as Code Size and Task Phase, were likewise designated as fixed effects. 

Random effects, conversely, account for unmanipulated sources of variation inherent in the sample population. Participant ID and Code Snippet ID were designated as random intercepts to control for individual differences between participants and difficulty variances across specific coding tasks. To determine final model configurations and avoid over-parameterization, we utilized Akaike Information Criteria (AIC) model comparison. AIC balances goodness-of-fit against model complexity by penalizing additional parameters, guarding against overfitting (where a model captures sample noise rather than true underlying patterns)~\cite{chakrabarti_aic_2011}. Because maximal LME specifications risk convergence failures, we started from a baseline specification (the primary effect of interest and random participant intercept) before sequentially evaluating additional variables, retaining them only if they lowered the AIC score. For instance, to evaluate whether broader student demographics (Gender, GPA, and LLM Usage Frequency) introduced meaningful baseline variance, we tested them in the model; however they explained negligible variance and increased AIC scores, threatening convergence without improving fit. Consequently, they were omitted. In contrast, Participant ID, Snippet ID, Code Size and Task Phase were retained because their inclusion consistently optimized model performance. This systematic procedure prevented arbitrary feature selection while accurately accounting for the repeated-measures dependencies of the data prior to conducting likelihood ratio tests.

With the final model structures established, hypothesis testing was executed via Likelihood Ratio Tests (LRT)~\cite{bates_fitting_2015}. An LRT evaluates a specific factor by comparing a "full" model containing the term of interest (such as AI assistance) against an identical null model lacking only that term. Because each test must isolate a single factor or interaction, testing multiple effects requires separate model comparisons; bundling multiple new variables into a single test would obscure which specific variable is driving any observed change in model fit. If the LRT shows that the full model fits the data significantly better than the null model, it means the factor in question has a statistically significant impact. 

To isolate specific facets of how AI-assistance impacted cognitive workload across our tDRT and questionnaire (NASA-TLX \& SEQ) measurements, our analysis was partitioned into six LME models and corresponding LRTs (summarized in Table~\ref{tab:LME_model_results}):
\begin{itemize}
    \item Model~1 (Global Main Effect): Evaluates the overall impact of AI-assistance on tDRT reaction times across the entire study.
    \item Model~2 (Task Phase Interaction): Tests whether the cognitive impact of AI-assistance varied significantly between code reading phase and quiz phase.
    \item Model~3 (Code Size Interaction): Tests if the impact of AI-assistance changed based on the size of the code snippet.
    \item Model~4 (Java Experience Interaction): Examines whether a participant's programming experience affects how AI-assistance impacted their reaction time.
    \item Model~5 \&~6 (Subjective Workload): Evaluates the impact of AI-assistance on perceived workload and task difficulty, using unweighted NASA-TLX sub-scale averages and single-item SEQ responses respectively. 
\end{itemize} 

%retaining random intercepts for Task ID and by-participant random slopes for the AI-assistance
%Following AIC optimization, and to ensure consistency across all six likelihood ratio tests, the same random effects structure was maintained for each model, accounting for random intercepts by code snippet and by-participant random slopes.

\subsubsection{RQ2 Data Analysis}
Code comprehension was assessed via the accuracy score ($0$--$4$) on the four multiple-choice questions per task, analyzed in R using the same \texttt{lme4}/\texttt{lmerTest} toolchain as RQ1. We first computed descriptive accuracy means by AI condition and code size (Table~\ref{tab:comprehension_results}). Our planned analysis modeled the $0$--$4$ accuracy score using a linear mixed-effects model with AI assistance as a fixed effect and by-participant and by-snippet random intercepts, testing the effect of AI assistance via a likelihood ratio test as in RQ1. 
Because a four-point sum is a bounded, discrete outcome that does not fully satisfy the assumptions of a linear model, we additionally fit an ordinal (cumulative-link) mixed model with the same fixed- and random-effects structure as a sensitivity analysis, using the \texttt{ordinal} package~\cite{christensen_ordinal}. 

\subsubsection{RQ3 Data Analysis}
We analyzed participants' responses to the two open-ended questions on trust and reading-behavior changes using inductive thematic analysis. Responses were first segmented into discrete meaning units to isolate distinct ideas. Two researchers collaboratively conducted open-coding on these segments, organizing them into an affinity diagram using Miro~\cite{miro_software}. %\footnote{https://miro.com/}. 
Affinity diagramming is a common synthesis technique in UX research that supports the collaborative grouping of related observations into higher-level themes~\cite{holtzblatt_rapid_2004}. Through a negotiated agreement strategy~\cite{garrison_revisiting_2006}, a third researcher joined discussions to iteratively refine groupings into coherent themes, which were then reviewed by a fourth researcher with expertise in qualitative analysis and human factors in software engineering. No predefined codebook was used; instead, themes were developed inductively from the data.

\subsection{Study Procedures}
Upon arriving at the study location, participants reviewed an informed consent form and provided verbal consent, with the reminder that they could withdraw at any time. The experimenter then briefed them on the overall procedure, the code tasks, and the use of the tDRT. The participant was then fitted with the tDRT equipment, placing the response button on the thumb of their non-dominant hand and securing the vibrating motor secured to their shoulder or arm (Figure~\ref{fig:DRT_set_up}). To ensure familiarity with the tDRT stimulus and response mechanism, participants completed a brief practice session with a sample code snippet and a single multiple-choice question, before continuing to the primary experimental tasks.

\begin{figure}[h]
    \centering
    \includegraphics[width=1\linewidth]{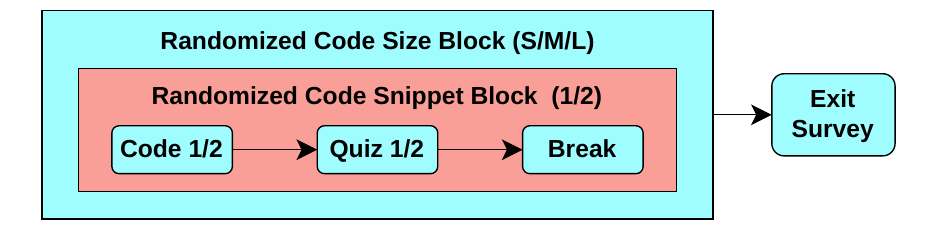}
    \caption{Study Procedure Diagram}
    \label{fig:Study Procedure Diagram}
\end{figure}

\begin{figure}[h]
    \centering
    \includegraphics[width=0.8\linewidth]{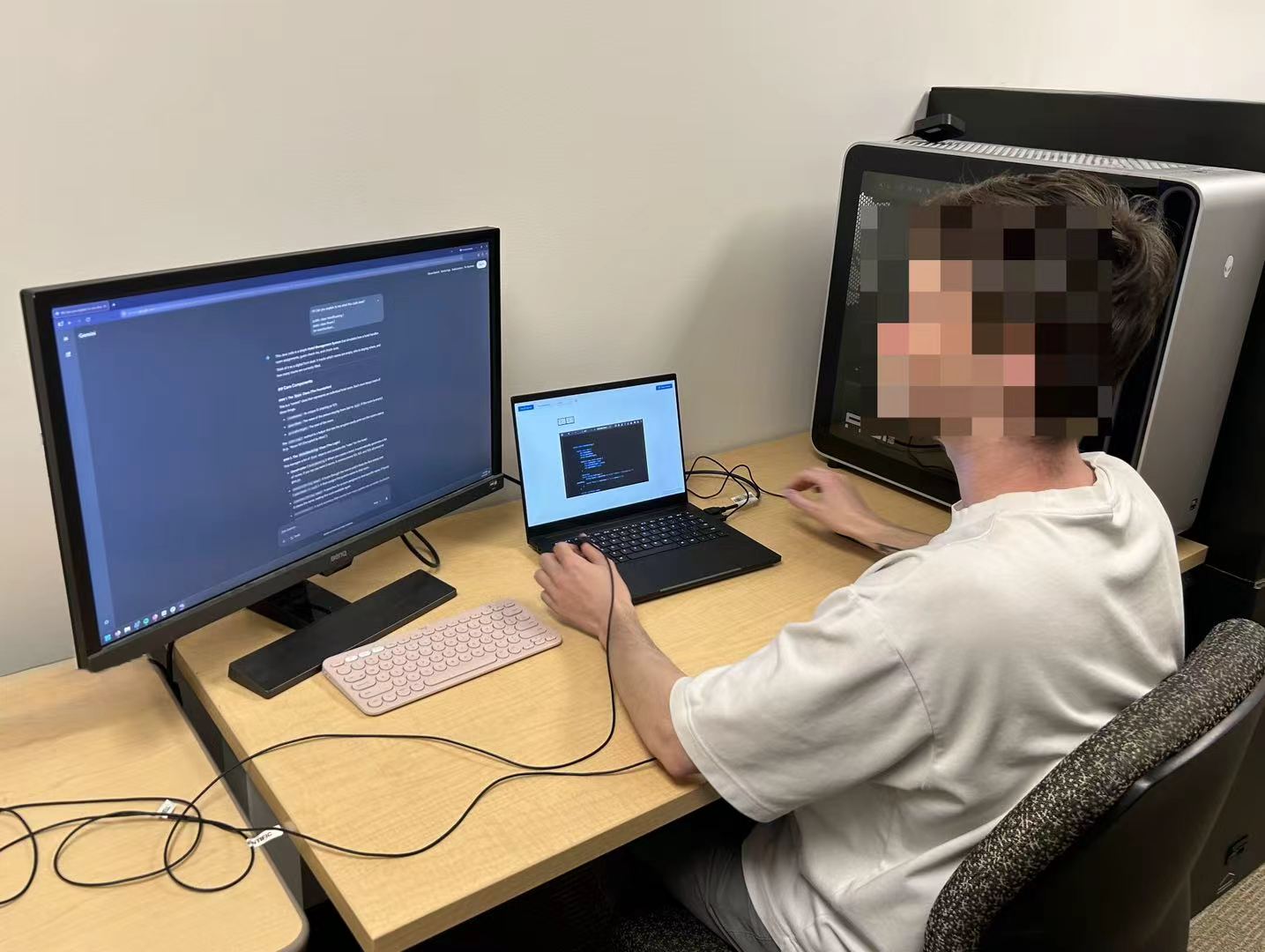}
    \caption{Study Set Up}
    \label{fig:study_table_setup}
\end{figure}

Tasks were completed at a workstation set-up with a laptop and adjacent secondary monitor. Code snippets and quiz questions were displayed on the laptop, while the connected monitor was reserved for displaying the Gemini browser interface during the AI-assisted code reading (Figure~\ref{fig:study_table_setup}). The presentation order of the three code size categories and the task order within each size category were randomized, as detailed in Figure~\ref{fig:Study Procedure Diagram}.

\begin{table}[htb]
    \centering
    \caption{Study Measurements}
    \small
    \begin{tabular}{lcl}
        \toprule
        \textbf{Measurement} & \textbf{Variable Type} & \textbf{Source} \\ 
        \midrule
        tDRT reaction times & Dependent & tDRT Device~\cite{noauthor_iso_nodate} \\ 
        Quiz scores & Dependent & Comprehension Quiz~\cite{zenodoreppackageDraft} \\ 
        TLX and SEQ scores & Dependent & Post Quiz Survey~\cite{doi:10.1177/154193120605000909,sauro_chapter_2012} \\ 
        Background questions & Independent & Post Study Survey~\cite{zenodoreppackageDraft} \\ 
        \bottomrule
    \end{tabular}
    \label{tab:measurements_clean}
\end{table}

\begin{table}[htbp]
\centering
\caption{Experimental Protocol for Participant \#1}
\label{tab:p1_protocol}
\begin{tabular}{@{}ccc@{}}
\toprule
\textbf{Task Order} & \textbf{Code Size \& ID} & \textbf{AI Assist.} \\ \midrule
1                   & Small1              & No                      \\
2                   & Small2              & Yes                     \\
3                   & Medium2             & Yes                     \\
4                   & Medium1             & No                      \\
5                   & Large2              & Yes                     \\ 6                   & Large1              & No                      \\
 \bottomrule
\end{tabular}
%}
\end{table}

Each of the six tasks followed a three-phase sequence: 
\begin{enumerate}
    \item Code Reading Phase: Participants analyzed the code snippet with free access to the secondary LLM window (AI-assisted condition), or with the screen minimized (control condition). To ensure sufficient engagement with the code, a "Next" button only appeared after a size-dependent minimum threshold (2.5, 3, or 3.5 minutes). Participants automatically advanced to the next phase if the maximum time limit was reached (4, 5, or 6 minutes).
    \item Comprehension Phase: Participants completed the four multiple-choice questions on the laptop screen. Participants could manually advance after answering all questions, or were automatically advanced if a maximum time limit was reached (2, 2.5, or 3 minutes).
    \item Subjective Evaluation Phase: The tDRT device was deactivated, and participants completed a NASA-TLX questionnaire and SEQ to report perceived workload and task difficulty. 
\end{enumerate} 
Following the questionnaire, participants rested for one minute before proceeding to the next task. 

% This slightly variable time window was implemented to ensure sufficient engagement with the code material, while preventing boredom or frustration due to varying reading or comprehension speeds.

\section{Analyses and Results}
\label{sec:results}

% \textit{The results section should contain the results you obtained by executing the steps described in the methodology. Additionally, it should contain an answer to each research question.}
Twenty-two participants completed the study, but data from two were excluded from analysis: one due to technical issues that disrupted the study structure, and one who did not interact with the AI system during the AI-assisted condition, rendering their AI-condition data invalid. The resulting twenty participants' demographic information  is presented in Table~\ref{tab:demographics}.

We now address the research questions introduced in Section~\ref{sec:introduction}.

\subsection{How does access to GenAI assistance affect developers' real-time cognitive workload during code comprehension? (RQ1)}
\label{sec:results-rq1}

RQ1 is answered by assessing developer workload through objective tDRT response times and subjective NASA-TLX and SEQ metrics. As detailed in the methodology(Section~\ref{subsec:DRT_methodology}), raw tDRT data underwent pre-processing to exclude typing intervals and invalid reaction times, yielding a final dataset of $8,048$ tDRT responses for statistical modeling. To isolate the specific facets of this data we executed six LME models, with four analyzing the tDRT metrics and two evaluating the questionnaires, as summarized in Table~\ref{tab:LME_model_results}.

% \begin{table}[htbp]
%     \centering
%     \caption{Fixed and Random Effects}
%     \begin{tabular}{c|c} 
%     \toprule
%     \textbf{Fixed Effects}  & \textbf{Random Effects}  \\
%     \midrule
%         AI Condition & Participant ID \\
%         Code Size   & Task ID \\
%         Task Phase &  \\
%         Java Experience & \\
%         %LLM Familiarity & \\
%     \bottomrule
%     \end{tabular}
%     \label{tab:fixed_random_effects}
% \end{table}

\begin{table}[htbp]
    \centering
    \caption{tDRT \& Questionnaire LRT Results}
    \label{tab:LME_model_results}
    \begin{tabular}{l|c|c|c|c|c}
    \toprule
      \textbf{Model} & \textbf{DV} &\textbf{Tested Effect}&
      \textbf{$\chi^2$} & \textbf{DF} & \textbf{p-value} \\
        \midrule
        
        Model 1 & RT &AI Assistance & $17.37$ & $1$ & $<.001^{**}$ \\
        \hline
        
        Model 2 & RT & AI$\times$Task Phase & $7.10$ &$1$ & $0.008^{**}$ \\
        \hline
        
        Model 3 & RT & AI$\times$Code Size & $8.22$ &$2$ & $0.016^{*}$  \\
        \hline
        
        Model 4 & RT & AI$\times$Java Exp & $0.89$ &$3$ & $0.828$  \\
        \hline
        
        Model 5& TLX Score& AI Assistance & $1.80$&$1$ & $0.180$ \\
        \hline
        
        Model 6& SEQ Score& AI Assistance& $1.96$&$1$ &$0.161$ \\
    \bottomrule
    \end{tabular}
    \vspace{1ex}
    \begin{flushleft}
    \footnotesize 
    \textit{Note:} $^{*}$Statistically significant at the $p <.05$ level; $^{**}$significant at the $p < .01$ level.
    \end{flushleft}
\end{table}

\begin{figure}
    \centering
    \includegraphics[width=0.8\linewidth]{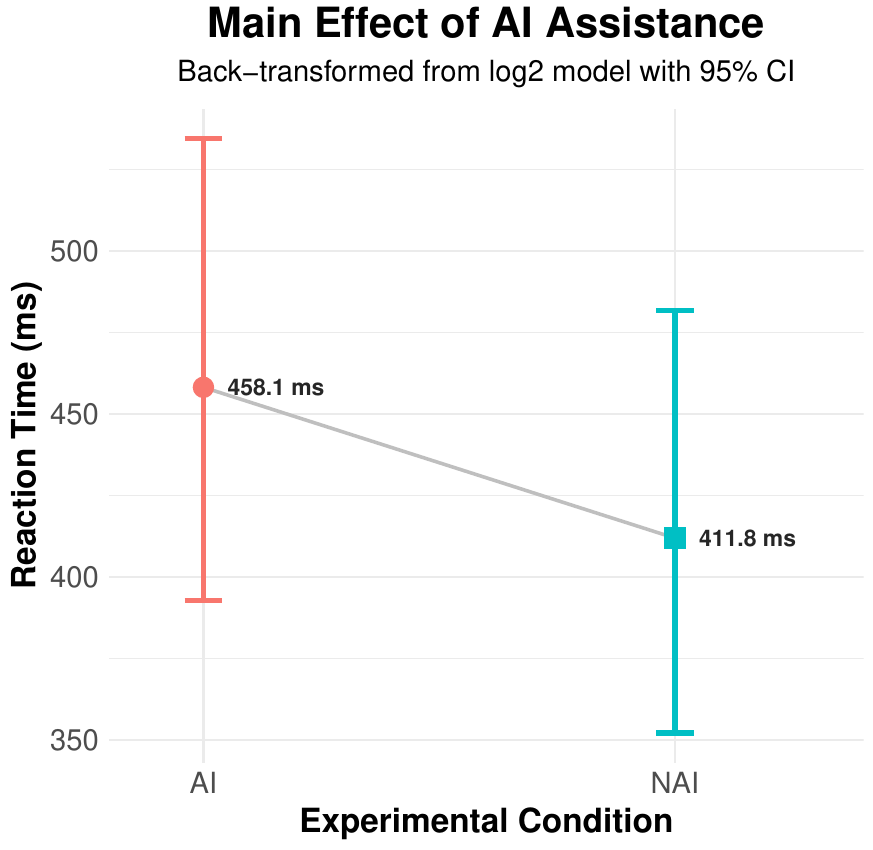}
    \caption{Main Effect of AI Assistance on Cognitive Workload}
    %\\ Error bars represent 95\% confidence intervals back-transformed from the $\log_2$ model. The AI condition resulted in a significant 11.2\% increase in reaction time compared to the No-AI baseline ($p < .001$)}
    \label{fig:ai_DRT_response}
\end{figure}

Model 1 isolated the global main effect of AI assistance on tDRT reaction times, specifying AI Assistance, Task Phase, and Code Size as fixed factors. A likelihood ratio test confirmed that the inclusion of the AI Assistance factor significantly improved overall model fit ($\chi^2 = 17.37$, $p < .001$, see Table~\ref{tab:LME_model_results}).
This indicates that \textbf{the presence of AI assistance was a significant predictor of reaction time, which indicates an increase in cognitive workload}. Back-transforming the model coefficients from the $\log_2$ scale revealed that AI assistance increased average reaction times by $11.2\%$. This effect is visualized in Figure~\ref{fig:ai_DRT_response}, which displays the estimated marginal means of tDRT response times across conditions.

To assess whether the impact of AI assistance varied by experiment stage (reading or quiz), we evaluated the interaction between AI Assistance and Task Phase (Model 2, Table~\ref{tab:LME_model_results}). The LRT confirmed a significant interaction ($\chi^2=7.10$, $p = 0.008$), indicating that the influence of AI on cognitive workload was not uniform across task stages. 

\begin{figure}
    \centering
    \includegraphics[width=0.8\linewidth]{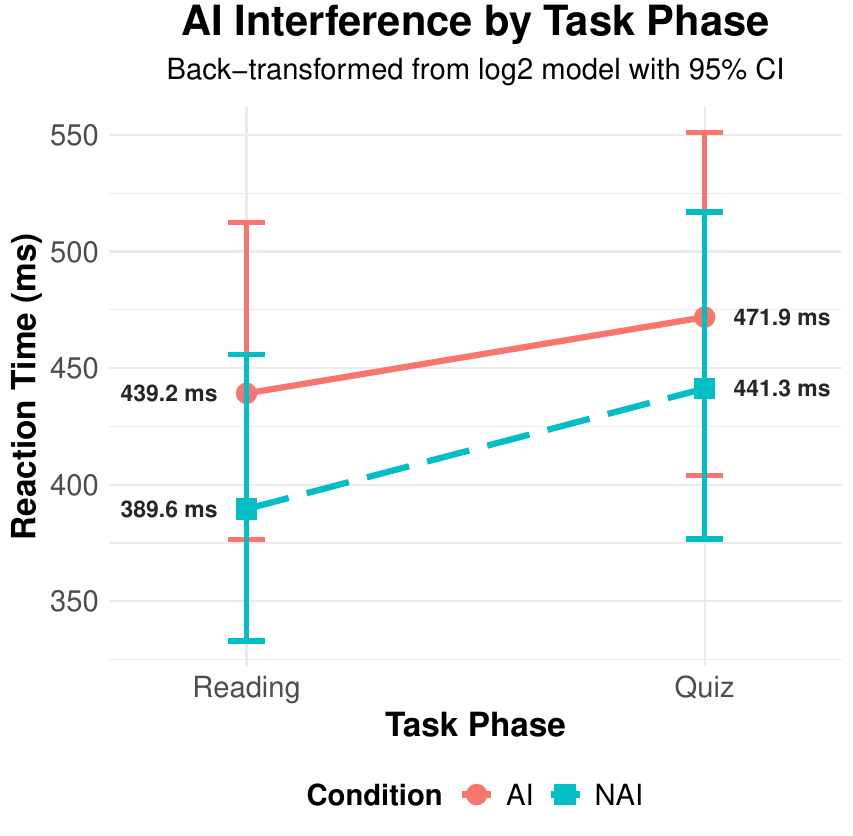}
    \caption{Interaction between AI Assistance and Task Phase on Cognitive Workload}
    \label{fig:ai_rq_interaction}
\end{figure}

As illustrated in Figure~\ref{fig:ai_rq_interaction}, overall reaction times increased for both groups when moving to the quiz phase. The control condition exhibited a steeper relative increase across this transition, which was due to starting from a lower baseline during the reading phase. Specifically, post-hoc pairwise comparison showed that during code reading, AI-assistance elevated reaction times by $12.7\%$~($p<.001$). Once participants moved to the quiz phase, this gap narrowed to a smaller, but still statistically distinct $6.9\%$~($p=0.007$) increase compared to the control condition. These results demonstrate a significant carryover effect, where \textbf{AI usage during the reading phase resulted in a higher cognitive workload during the subsequent quiz phase, despite AI no longer being in-use}. 

Model 3 evaluated the interaction between AI-assistance and Code Size to determine if the impact of AI on cognitive workload changed based on code snippet size. The LRT confirmed a significant interaction effect ($p=0.016$), though AI assistance elevated reaction times across all three size categories: Small (8.9\%, $p<.001$), Medium (15.2\%, $p<.001$), and Large (9.5\%, $p<.001$). 

\begin{figure}
    \centering
    \includegraphics[width=0.8\linewidth]{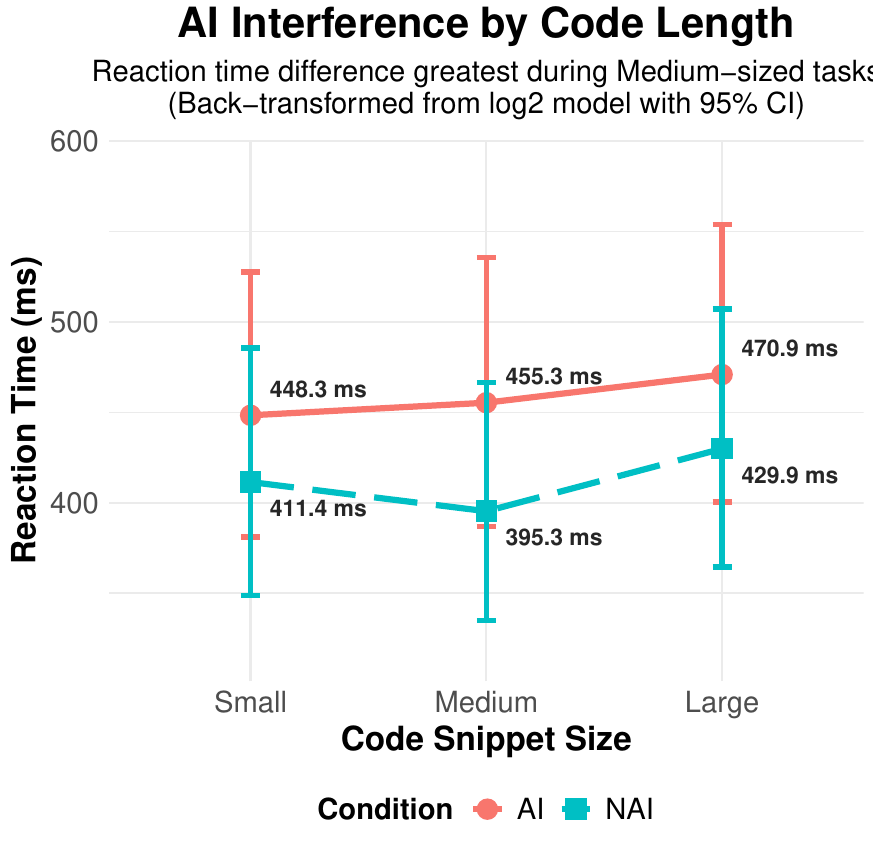}
    \caption{Interaction between AI Assistance and Code Size on Cognitive Workload}
    \label{fig:ai_size_interaction}
\end{figure}

Post-hoc interaction contrasts indicated that the magnitude of this reaction time increase was significantly more pronounced in medium-sized tasks than in either small (z = 2.47, p = 0.036) or large tasks (z = 2.44, p = 0.039). The contrasts also confirmed that the AI-influenced reaction time increases in the small and large tasks were statistically equivalent to each other (z = 0.25, p = 0.968). Interestingly, Figure~\ref{fig:ai_size_interaction} shows that this variance is caused by a baseline dip within the control condition during medium-sized tasks, rather than a spike in workload caused by AI usage. Ultimately, \textbf{AI-assistance overrode the significant relief in cognitive load observed in the control condition during medium-sized tasks, maintaining a persistently higher workload across all code lengths}. 

Finally, Model 4 evaluated the interaction between AI Assistance and Java Experience to assess whether the impact of AI on cognitive workload varied with user expertise (Table~\ref{tab:LME_model_results}). No significant interaction was found between these variables ($p = 0.828$), demonstrating that the \textbf{AI-induced cognitive workload increase remained consistent regardless of participants' level of Java familiarity}.

Following the tDRT analysis, Models 5 and 6 evaluated the questionnaire data to determine if the presence of AI-assistance altered participants' perception of their workload and task difficulty (Table~\ref{tab:LME_model_results}). The LRTs showed that AI-assistance did not significantly impact scores for either the NASA-TLX ($p=0.180$) or the SEQ ($p=0.161$). These results indicate that \textbf{while AI assistance significantly increased objectively measured cognitive workload, it did not lead to a statistically significant change in perceived workload or self-reported task difficulty}.

\MyBox{\textbf{RQ1.} AI assistance significantly increased objective cognitive workload, as indicated by tDRT reaction times 11.2\% higher in the AI-assisted condition. This effect persisted beyond active tool use: even after AI was removed, reaction times in the AI-assisted condition remained elevated during the quiz phase. Perceived workload, as measured by NASA-TLX and SEQ, was unchanged between conditions. Rather than reducing cognitive load, AI assistance introduced a dual-task penalty that participants did not detect through
self-report alone.}

\subsection{How does access to GenAI assistance affect observable code comprehension performance across tasks of varying complexity? (RQ2)}

To evaluate whether access to generative AI improved code comprehension, we analyzed accuracy on the four-question quiz (scored $0$--$4$) using our planned mixed-effects model, with AI assistance as a fixed effect and by-participant and by-snippet random intercepts.

\begin{table}[htbp]
\centering
\caption{Code Comprehension Performance by Condition}
\label{tab:comprehension_results}
\begin{tabular}{@{}lcc@{}}
\toprule
\textbf{Measure} & \textbf{AI Assisted} & \textbf{No AI} \\ \midrule
Mean accuracy, 0--4 (SD) & 3.22 (0.99) & 3.23 (1.08) \\
Complete comprehension (all 4 facets)$^{*}$ & 50.0\% & 58.3\% \\
\bottomrule
\end{tabular}
\vspace{1ex}
\raggedright \footnotesize \textit{$^{*}$Descriptive rate of tasks receiving a perfect score; reported to characterize the ceiling effect.}
\end{table}

Mean accuracy was high and nearly identical across conditions (AI: M$=3.22$, SD$=0.99$; No-AI: M$=3.23$, SD$=1.08$; Table~\ref{tab:comprehension_results}). Scores clustered near the top of the scale: a perfect score across all four facets was achieved on roughly half of all tasks. Because the four question types award partial credit across distinct comprehension facets (maintenance, summarization, mental execution, structural mapping)~\cite{10.1145/3626522}, participants reliably answered most individual facets, compressing the aggregate score toward its maximum and producing a ceiling effect.

The planned model found no significant effect of AI assistance on comprehension accuracy ($\chi^2(1)<0.01$, $p=.982$). The ordinal sensitivity analysis agreed, likewise showing no significant effect of AI ($\beta=-0.16$, $z=0.40$, $p=.692$), confirming the null was not an artifact of fitting a linear model to the bounded, ceiling-affected scale.

We therefore found no evidence that AI assistance improved code comprehension at the difficulty levels evaluated in this study. We read this as an absence of evidence for a benefit, not as evidence that AI confers none.

\MyBox{\textbf{RQ2.} AI assistance did not improve code comprehension. Quiz scores were high and nearly identical between conditions (AI: M$=3.22$; No-AI: M$=3.23$), and both the planned mixed-effects model ($p=.982$) and an ordinal sensitivity analysis ($p=.692$) found no significant effect of AI. Because scores clustered near the maximum, the study had limited power to detect benefits on harder tasks: we find no evidence that AI improved comprehension at the difficulty levels tested, rather than evidence that it provides none.}

\subsection{How do developers perceive the impact of GenAI assistance on their overall code comprehension process and strategies? (RQ3)}
\label{sec:results_rq3}

    % Within our qualitative analysis on the impact of GenAI assistance, we extracted codes from participant open-text responses. These responses were mapped along a two-dimensional spectrum ranging from tool-utilization (\textit{No-Use} to \textit{High AI-Use}) and user autonomy (\textit{Human Agency} to \textit{Over-Reliance}). Within this spectrum, participant strategies coalesced into three distinct themes that reflect the interplay between their trust and reading behaviors. 
% ^ Re-framed the diagram to emerging themes rather than establishing a new model:

To understand the human factors underlying our objective metrics, we analyzed the open-text post-task responses regarding both participants' trust in the LLM and its impact on their code reading behavior. Our thematic analysis of these responses revealed that the trust and reading behaviors were intersectional, the level of trust impacted how they utilized the AI, and the output received during usage impacted their trust. As such, we utilized an affinity diagram to analyze response snippets together. From this combined dataset, we identified three behavioral themes: 

%how participants trusted the output provided by the LLM during the code comprehension tasks and found themes related to xxxxx.
%The analysis about whether and how the AI-assistance had changed their approach to reading code and found 
%themes related to different \textsc{behaviors} about AI reliance, resistance, scaffolding and displacement.

%, we coded each statement into categories, most prominently \textit{Trust}, \textit{LLM Output}, and \textit{Behavior}. In doing so, we noticed that responses varied along two informal spectrums: how heavily a participant used the AI (\textit{No-Use} to \textit{High AI-Use}), and how far they relied on it over their own judgement (\textit{Human Agency} to \textit{Over-Reliance}). This split was especially visible among heavy users: some checked the AI's every claim against their own reading of the code, while others scarcely looked at the code and accepted its explanations wholesale. As an informal aid to make sense of these patterns, we arranged coded statements that fit into these spectrums onto a simple two-dimensional plot defined by the two spectrums, letting us gauge roughly how many fell into each region. Because a single participant's response sometimes specified more than one distinct behavior, we split such responses into separate statements, resulting in 26 fitting statements in total across the 20 participants. This exercise surfaced three recurring themes in the interplay between trust and reading behavior, which we term \textit{Resistance}, \textit{Scaffolding}, and \textit{Displacement}.

\textbf{Resistance (Low/No AI-Use, High Human Agency)}
Participants exhibiting resistance
%(9 of 26 coded statements) consciously minimized their reliance on the GenAI,
reported using the GenAI only to confirm an understanding they had already formed independently. Within this strategy, trust was restricted, with the AI utilized only for validating higher-level structures or independently verified conclusions: \textit{``I trust it enough to get me the bones of a piece of useful code, but not the full thing.''}
 %Other quote option: ``No, because in order to understand what the code is doing I need to manually look it over and follow the path.''
 For these individuals, code reading mechanics remained largely unchanged: \textit{``I still read the code top down''}

\textbf{Scaffolding (Moderate/High AI-Use, Balanced Human Agency)}
The most common strategy was scaffolding,
%(11 of 26 coded statements)
where participants used the GenAI as an initial orientation tool to provide a structural overview that accompanied their reading of the code. Within this theme trust was conditional, varying according to how complex a participant perceived the task to be. Where the code felt straightforward, participants accepted the AI explanation at face value: \textit{``I generally trusted it because the code was relatively simple.''}

Conversely, for responses addressing more complex logic, participants maintained agency by verifying the AI against their own developing mental model: \textit{``I fully trusted it only if what I though[t] was the output matched the LLM's interpretation and I was also semi-confident in my solo interpretation.''} This scaffolding approach reshaped participants' reading mechanics, allowing them to bypass low-level syntax tracking in favor of a macro-level evaluation of the code: \textit{``It moved me from reading the code in smaller blocks to reading it in larger blocks by method.''}

\textbf{Displacement (High AI-Use, High Over-Reliance)}
The third identified strategy was displacement,
%(6 of 26 coded statements)
where participants allowed the GenAI output to actively substitute for direct code reading. Within this theme, trust shifted towards over-reliance, causing participants to minimize or entirely bypass linear code analysis. This resulted in participants lightly skimming the text or offloading the cognitive task of comprehension to the AI: \textit{``I then almost exclusively read the LLM output instead of the code.''} For some the code was reduced to a secondary reference: \textit{``[I ended up] only glancing at the code when I had a small question or curiosity about the way it was implemented.''} This substitution did not always pay off; one participant reported it cost them comprehension, as they \textit{``took longer to read the excessive and incomprehensive writing by the ai and missed a lot of vital information that just reading the code would have imparted.''} %The long-term implications of this displacement shift are discussed in Section~\ref{sec:discussion}

    % \MyBox{\textbf{RQ3.}
    % Participant comprehension strategies map along a spectrum of autonomy and tool utilization, revealing that trust in GenAI was conditional and mediated by task-complexity. While many leveraged the tool as a macro-level reading guide (Scaffolding), 20\% allowed the natural-language explanations to displace linear code analysis entirely (Displacement)}
\MyBox{\textbf{RQ3.} Coding participants' open-text responses surfaced three recurring strategies, distinguished by how much they used the AI and how far they relied on it, with trust consistently conditional and scaled to perceived task complexity. Some set the tool aside and read independently (\textit{Resistance}); most used it as a macro-level reading guide while checking it against the code (\textit{Scaffolding}); and roughly a quarter let its explanations displace direct code reading entirely (\textit{Displacement}).}

\section{Discussion, Limitations and Implications}
\label{sec:discussion}

\textbf{A comprehension aid that adds load.} Prior work has framed GenAI assistance primarily as a tool for reducing the cognitive burden of code comprehension, offloading explanation and summarization to the model~\cite{nam_using_2024}. Our results
challenge that framing. AI assistance did not reduce workload; it increased it, reliably and across all code sizes and experience levels. The reason, we argue, is that AI introduces a dual-task structure that prior framing overlooked. The comprehension task does not disappear when AI is present; rather, it is joined by a second task of managing the AI interaction. This mental overhead is explicitly reflected in our qualitative findings, where participants across the \textit{Resistance} and \textit{Scaffolding} themes reported consciously verifying the AI output against the code. This active verification behavior embodies the precise mechanism of our proposed dual-task load, where the developer must simultaneously process the code and compare the AI's output against their own mental model. Whether the net effect is ultimately beneficial depends on whether the assistance gained outweighs the overhead incurred, a trade-off future work with more complex tasks should prioritize. 
%, a question our data cannot fully answer given the ceiling effect in performance scores, but one that future work with harder tasks and broader populations should prioritize.

\textbf{Feeling easier does not mean working less.} Despite an 11.2\% elevation in tDRT reaction times, neither NASA-TLX nor SEQ scores differed between conditions. This dissociation between objective and perceived workload is not new~\cite{yeh1988dissociation, fritz2014using}, but GenAI creates a specific condition that prior work did not anticipate. Unlike a complex query language or an unfamiliar IDE, which feel demanding, GenAI assistants are designed to reduce apparent friction. Formulating prompts, evaluating text-heavy model replies~\cite{catalan2026m}, and switching attention between code and AI interface~\cite{trakoolgerntong2025ailinkpreviewer} all consume attentional resources beneath awareness, producing a metacognitive blind spot that self-report instruments like the NASA-TLX are structurally unable to
detect~\cite{yeh1988dissociation, fritz2014using}.

\textbf{The cost that stays when the tool leaves.} 
Our results show that once the AI was removed, participants in the AI-assisted condition still exhibited reaction times 6.93\% higher than controls
($p = 0.007$) under identical conditions. If the cost of using AI were simply tool-operation overhead, it would have disappeared when the tool was removed. Instead, this sustained elevation indicates that AI assistance likely short-circuited the active inference and execution tracing %required for the construction of a working mental model of the code.
necessary to build a stable mental model. When the tool was removed during testing, participants had a weaker cognitive foundation, making the quiz phase significantly more effortful. This behavior appears to extend traditional automation complacency
frameworks~\cite{parasuraman2010complacency}, suggesting that the primary risk of generative tools may not be merely over-reliance, but a displacement of the effortful processing required for independent knowledge construction -- a hypothesis with implications for long-term encoding and recall that future longitudinal research should directly test.
%: the primary risk of generative tools is not merely over-reliance, but a displacement of the effortful processing required for long-term knowledge encoding and independent recall.

\textbf{The overhead is in the tool, not the task.} The interaction analyses revealed that developer experience
did not moderate the workload increase. Participants with more Java familiarity were more efficient readers in the
unassisted condition, yet this efficiency did not reduce the attentional cost of AI interaction, as the AI Assistance $\times$ Java Experience interaction was not significant (see Table~\ref{tab:LME_model_results}, Model 4, $p = 0.828$). Although the AI Assistance $\times$ Code Size interaction was significant (see Table~\ref{tab:LME_model_results}, Model 3, $p = 0.016$), post-hoc contrasts revealed that the AI-induced elevation was statistically equivalent at small and large sizes (8.9\% and 9.5\% respectively;
$z = 0.25$, $p = 0.968$), and the apparent medium-size difference was driven by a control-condition dip rather than a spike in AI-induced workload. If AI were reducing
intrinsic load, its overhead should shrink where code is simpler and expand where code is harder, yet the AI-induced cost remained largely constant across sizes. Future research should investigate whether this pattern holds under conditions of greater complexity and wider experience ranges than those studied here.

\subsection{Threats to Validity and Limitations}
\label{sec:threats}

\subsubsection{External Validity.}
Several aspects may limit the generalizability of our findings to the broader software engineering community. Most notable, our sample was recruited entirely from a university student population. The cognitive workload profiles of students, who are frequently still consolidating syntax and architectural structures, may not generalize to professional software engineers with established mental models. Furthermore, because these student participants held varying levels of GenAI familiarity and were still developing their interaction strategies, our results likely capture the cognitive load of relatively novice AI adoption, rather than mature, expert workflows. Moreover, the AI agent was accessed via a web interface on a secondary monitor to maintain experimental control. In professional settings, generative tools are often integrated directly into the IDE. The contextual switching overhead imposed by an external browser interface may differ from the workload associated with native integrations~\cite{haque_towards_2025, nam_using_2024}.

\subsubsection{Internal Validity.}
A primary threat to our internal validity involves potential mechanical interference during the AI-assisted reading phase, where the physical act of typing prompts could disrupt responses with the finger-mounted tDRT button. To isolate cognitive workload from these physical artifacts, all trials involving active typing were removed from our analysis. Furthermore, the highest workload measurements occurred during the subsequent quiz phases where no typing took place, confirming that mechanical interference from tool interaction does not account for the sustained workload impact. 

%Second, participants were not provided specific instructions or training on prompting strategies for code comprehension. As such, a portion of the recorded cognitive load during the reading phase may stem from participants unfamiliar with the tool learning to use it, rather than the comprehension task itself. 

%\subsubsection{Construct validity.}
%A potential threat to our construct validity is the artificial nature of the experimental task. Instructing participants to use an AI agent for code comprehension may not reflect the natural process developers would follow in real-world settings. Participants who do not typically utilize LLMs may have experienced extraneous cognitive load related to managing and querying the interface, rather than workload from the comprehension task itself. 

\subsubsection{Conclusion validity.}
The main threat to our conclusion validity is that elevated cognitive workload is not inherently detrimental; it can sometimes reflect higher engagement that leads to better information retention. 
While we believe the elevated cognitive load observed in our study was due to dual-task demands, because of the ceiling effect in our performance data, we are limited in our ability to determine if the increased cognitive load would have yielded performance benefits in more complex programming tasks. This is an avenue for future work, which should utilize a greater variety of task types and difficulty to help map if this additional load benefits or harms comprehension. 

%Because participants achieved near-perfect scores across quizzes, we cannot definitively determine whether this AI-induced load was productive or disruptive. 
%However, these findings remain valuable: while generative tools are intended to lower developer workload, our objective metrics show they introduce a significant, hidden cognitive cost. 

\subsection{Implications for Practice}

While our findings are grounded in a student sample working with short code snippets, they offer preliminary insights relevant to practitioners and educators integrating generative AI into software development workflows. 
 
First, practitioners must remain cautious when tools make programming tasks simply ``feel'' easier. The discrepancy between our subjective perceptions and objective measurements (the 11.2\% workload increase) indicates that developers may operate under a false impression of reduced effort. This aligns with the efficiency illusions observed by~\cite{yu_efficiency-gain_2026}, where users predicted AI-assisted tasks to be significantly faster despite completion times remaining equivalent to unassisted work. While their study suggested that reduced subjective mental effort drives this temporal miscalibration, our findings demonstrate that even this impression of reduced effort can mask objective cognitive strain. Consequently, stakeholders evaluating AI tool efficiency should incorporate objective cognitive load and performance metrics rather than relying solely on subjective feedback, which can inadvertently mask mental fatigue.

Second, the sustained workload elevation observed when the tool was removed during the quiz phase suggests that relying on AI during initial comprehension carries a lingering cost when independent execution is required. This aligns with recent work on GenAI dependency and cognitive offloading, which indicates that AI assistance during learning can induce a false sense of competence during use, but compromise subsequent unassisted performance~\cite{lodge_artificial_2026, oecd_oecd_2026}. Our workload metrics show that when this assistance is abruptly withdrawn, transitioning back to unassisted problem-solving demands greater mental effort than working independently from the start. 
%which implies that the short-term fluency gained from AI-assistance leaves a cognitive debt. 
%Educators and students should recognize that transitioning from AI-assistance back to solo problem-solving appears to be more cognitively taxing than working independently from the start, likely because the AI hinders the deep encoding required for independent recall.  

\subsection{Implications for Researchers}

Our study introduces key methodological and theoretical contributions for future human-centric software engineering and human-AI interaction research.

First, our findings suggest that relying solely on subjective questionnaires is insufficient for capturing the full cognitive impact of AI tools. The contrast between our equivalent questionnaire ratings across conditions and the significant 11.2\% objective workload increase provides a clear example of the \textit{workload dissociation paradigm}~\cite{yeh1988dissociation}.  Because users routinely adapt to increased dual-task demands without conscious awareness of their mental resource expenditure~\cite{vidulich_cognitive_1988}, self-reported instruments alone fail to capture a tool's true cognitive impact. Future empirical software engineering studies should adopt multimodal assessment frameworks that combine subjective scales with real-time objective metrics to expose these hidden fluctuations in cognitive load. 
Second, to our knowledge, this study represents the first successful application of the Detection Response Task (DRT) within a software engineering context. Prior software comprehension research has relied on eye-tracking or EEG, which require complex calibration and are sensitive to noise. Our results show that the tDRT serves as an accessible and sensitive metric for capturing real-time fluctuations in developer mental load. 

%Finally, the finding that AI increases cognitive workload across both task phases, encourages closer investigation into the verification-comprehension trade off. Future human-AI interaction should investigate which specific interactive components contribute most to this load, and evaluate whether varying styles of AI assistance support or impede long-term mental model formation. 

\section{Conclusion}
\label{sec:conclusion}

This study presents an empirical evaluation of how GenAI impacts software developers during code comprehension, investigating its effects on real-time cognitive workload, observable performance, and comprehension strategies.

We conducted a controlled within-subjects experiment ($N=20$) utilizing a tactile Detection-Response Task (tDRT). Our findings challenge the assumption that GenAI inherently reduces developer effort. Instead, AI assistance significantly increased cognitive workload, raising tDRT reaction times by $11.2\%$, an elevation that was sustained into the unassisted testing phase. Despite this additional cognitive load, participants' subjective difficulty ratings remained similar between conditions. Our qualitative analysis contextualizes this dissociation through three behavioral themes: \textit{Resistance}, \textit{Scaffolding}, and \textit{Displacement}, which illustrate how distinct interaction styles and verification behaviors drive the mental overhead of human-AI collaboration.  

While a performance ceiling effect limited our ability to differentiate final accuracy outcomes, our findings clearly expose the hidden mental workload accompanying GenAI use. As software engineering increasingly integrates human-AI collaboration, research must look beyond immediate task completion to design paradigms that support active mental model formation, and to investigate whether the sustained cognitive costs observed here carry implications for long-term technical expertise.

\bibliographystyle{IEEEtranN}
\footnotesize{\bibliography{References}}

\end{document}